# Neutral Radical Molecules Ordered in Self-Assembled Monolayer Systems for Quantum Information Processing

A. Tamulis*, V. I. Tsifrinovich•, S. Tretiak+, G. P. Berman+, D. L. Allara⊗

*Institute of Theoretical Physics and Astronomy, Vilnius University, Vilnius, Lithuania
•IDS Department, Polytechnic University, Six Metrotech Center, Brooklyn, NY 11201, USA
+Theoretical Division, Los Alamos National Laboratory, Los Alamos, NM, 87544, USA
⊗Department of Chemistry and the Materials Research Institute,
Pennsylvania State University, University Park, PA 16802, USA

**Abstract**: Implementation of quantum information processing based on spatially localized electronic spins in stable molecular radicals is discussed. The necessary operating conditions for such molecules are formulated in self-assembled monolayer (SAM) systems. As a model system we start with 1, 3 -diketone types of neutral radicals. Using first principles quantum chemical calculations we prove that these molecules have the stable localized electron spin, which may represent a qubit in quantum information processing.

## 1. Introduction

It was suggested recently [1] that SAM systems could be used to create a macroscopic ensemble of quantum entangled 3-spin groups as a first step in quantum information processing. The spins of such a group could be connected by dipole-dipole interaction. Application of a nonuniform external magnetic field would allow selective excitation of every spin inside the group. The proper sequence of resonant electromagnetic pulses would then drive all spin groups into the 3-spin entangled state. In the suggested proposal [1] the spins were associated with a single unpaired electron spin of a neutral radical molecule in the SAM. One of the key elements of this strategy is the proper choice of molecules for experimental implementation of the proposal. Involved in this choice are four criteria for the chemical structure of these molecules:

1) A specific group or structural elements to provide self-organization characteristics;
2) A specific group to provide attachment of the molecules to a substrate;
3) An unpaired, spatially localized electronic spin representing an elementary qubit;
4) Strong non-compensated valence bonds, which are responsible for the unpaired electron spin, to provide the chemical stability of a qubit.

Modern quantum chemical methods provide powerful tools for theoretical modeling and analysis of molecular electronic structure and may be used to guide the synthetic effort [2, 3]. In particular, the small carbon-centered π-radical molecules (8-14 atoms) possessing a β-diketo structure have been investigated in the scope of Hartree-Fock (UHF) and unrestricted density functional theory (UDFT) methods [4]. The authors [4] calculated optimized geometry and hyperfine interactions in small  molecules without addressing their suitability for quantum information processing.

In this paper we report the results of quantum-chemical calculations of several specific radical molecules that provide a simple starting point for a larger consideration in a variety of simple small organic radical moieties of potential use in quantum computation. We use the UDFT approach to analyze molecular electronic structure for chemical stability and spatial spin localization. These simulations show that the suggested neutral radicals satisfy the constraints listed above and may serve as a start for designing molecules to be used in experiments targeting quantum computation with SAM structures.

## 2. Computational methodology

Exact quantum mechanical treatment of a large molecular system is intractable. The electronic structure methods routinely use Born-Oppenheimer approximation to decouple electron and nuclear motions which is fully justified for most cases [2, 3]. Subsequently one needs to solve the Schrodinger equation for the electronic system only, where nuclear positions enter as parameters. Quantum chemical methods provide practical recipes for approximate solutions for many-electron molecular systems [2, 3]. In particular, the Hartree-Fock approximation maps the complex many-body problem onto an effective one-electron problem in which electron-electron repulsion is treated in an average (mean field) way and assumes the simplest antisymmetric wavefunction for $N$-electron molecule, i.e. a single Slater determinant

$$\Psi = <\chi_1 \chi_2 ... \chi_N> \qquad (1)$$

Here $\chi_i(x)$ are the molecular orbitals (MO). Following Roothaan's procedure [5] they are expanded as linear combinations of localized atomic basis functions $\varphi_k(x)$ :

$$\chi_i = \sum_k \varphi_k(x) C_{ki} \qquad (2)$$

In the UHF approach $x$ refers to both coordinate and spin variables:

$$\varphi_k(x) = \phi_k(\vec{r})\xi(s) \qquad \text{where } \xi(s)=\alpha \text{ or } \xi(s)=\beta \qquad (3)$$

($\alpha$ and $\beta$ correspond to spin "up" and "down", respectively).

The HF secular eigenvalue equation is derived variationally by minimizing the energy with respect to the choice of MOs, i.e. coefficients $C_{ki}$

$$\bm{FC} = \bm{SCE} \qquad (4)$$

where $\bm{C}$ is the matrix of coefficients $C_{ki}$, $\bm{S}$ is the overlap matrix coming from non-orthogonality of atomic basis functions $\varphi_k(x)$, $E$ is the eigenvector of the respective MOs energies, and $F$ is a Fock operator (an effective Hamiltonian for one electron) which depends on the electronic density matrix $\rho_{ij}$ given by

$$\rho_{ij} = \sum_k^{occupied} C_{ki} C_{kj}^+ \qquad (5)$$

where the sum is running over the occupied molecular orbitals. Eq. (4) is nonlinear and usually is solved iteratively using self-consistent field procedure.

In spite of its simplicity, the HF approximation is not very accurate because it does not include electronic correlation effects [2, 3]. DFT makes it possible to treat these correlations by mapping complex many-electron problem into an effective mean field problem with the same energy which is the functional of the electronic density. As a result, in the DFT one solves the same one electron problem (Eqs. (1)-(5)) where the Fock operator is replaced by Kohn-Sham operator $h$ which is a functional of the electronic density. In principle this mapping is exact, but the functional is unknown. Extensive research, however, has formulated accurate functionals suitable for many complex cases [6, 7]. A detailed description of DFT, self-consistent field (SCF) procedure, expressions for the ground state SCF energy, and Fock and Kohn-Sham operators are readily available from quantum chemical textbooks [2, 3, 6, 7].

To summarize, for a given molecule, our calculations start from a trial geometry (Cartesian coordinates of the nuclei). Using UHF or UDFT approach and SCF procedure we obtain a molecular energy which depends on these coordinates parametrically. A subsequent standard geometry optimization procedure [2, 7] minimizes the energy with respect to the nuclei positions. Special care was taken to verify that the obtained optimal molecular structure is a global minimum in the phase space of the nuclear (3n-6, n being the number of atoms) degrees of freedom.

Once the optimal geometry of the radical is calculated we analyze its molecular electronic structure for spatial electronic spin localization and chemical stability. The spin density is defined as

$$\sum_i \int ds\, |\chi_i|^2 - \sum_j \int ds\, |\chi_j|^2 \qquad (6)$$

where indices $i$ and $j$ run over molecular orbitals with $\alpha$ and $\beta$ spins, respectively. This spatially distributed density is further condensed to individual atoms to associate a particular spin to each atom using electron population analysis. We use a Mulliken-type of analysis. Other schemes, however, are possible [9].

To analyze the stability of the non-compensated valence bonds responsible for the unpaired spin we finally calculate the bond order (the overlapping population) between the corresponding two atoms [2, 8]:

$$A = 2 \sum_{i,j,r} n_r C_{ri} C_{rj} S_{ij} \qquad (7)$$

Here $i$ and $j$ run over the basis functions of the first and second atoms, respectively, $r$ counts all basis function orbitals on these two atoms, $n_r$ is the r-th diagonal element of $CC^+$ (the occupation number of r-th basis function orbital), and $S_{ij}$ is the overlapping integral for atomic basis functions $i$ and $j$ ($S_{ij}=1$ for $i=j$).

To obtain accurate results, two factors need to be accounted for: i) the quality of the density functional and ii) the quality of the molecular orbitals (extension of the phase space for single-electron states). We choose the unrestricted Becke's 3 parameter exchange functional [10] with non local Lee-Yang-Parr electron correlation [11] (DFT UB3LYP model). Currently, the UB3LYP model is considered to be the most appropriate model to take into consideration electron correlations in large open-shell neutral radical molecules [12, 13]. To obtain accurate optimal molecular geometries, we use the 6-311G** basis set which includes (5D, 7F) polarized atomic orbitals (the standard tables [14] give the appropriate basis set description). To analyze the spatial electronic spin localization and stability at a relaxed molecule geometry, we subsequently use extended EPR-II basis set (also tabulated in Ref. [14]). The EPR-II basis set includes the re-optimized Huzinaga-Dunning double-zeta [15, 16] basis sets augmented with additional polarization functions and uncontracted in outer core - inner valence region and provides a good accuracy for modeling various molecular properties in organic radicals as shown by V. Barone [17].

## 3. Analysis of electronic structure in neutral radical molecules

We use the standard Gaussian 98 program suite [8] for all quantum-chemical calculations presented in this section, and the Molden program for visualization purposes [18].

Figure 1 shows the structure of a neutral radical molecule with a β-diketone structure that can satisfy the conditions formulated in the Introduction. The net spin of the molecule is normalized to one unit.

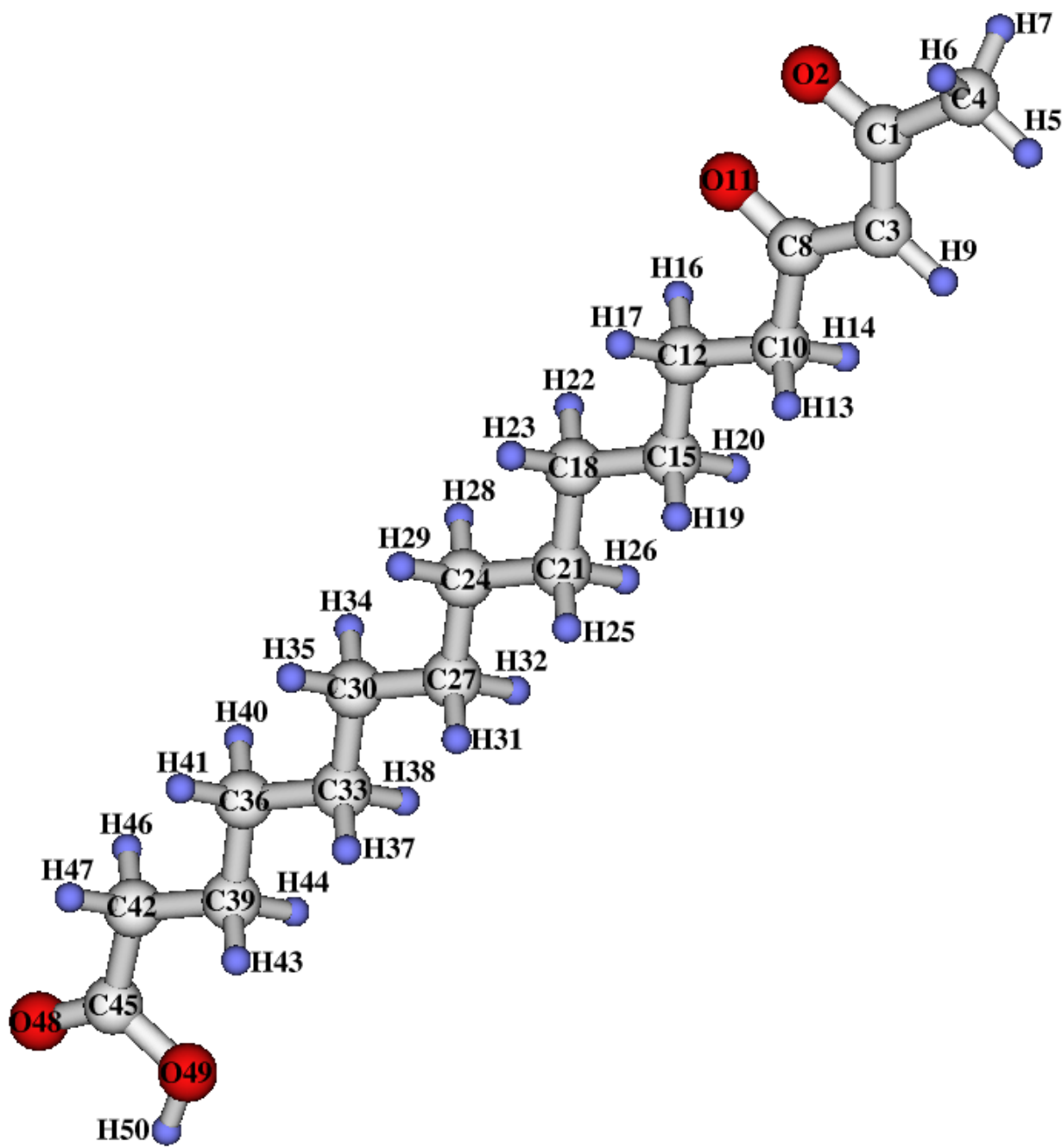

Figure 1. Geometry of optimized neutral radical molecule with β-diketone structure.

The molecule consists of 50 atoms (carbon, hydrogen, and oxygen) which are labeled according to Fig. 1. The group -(CH$_2$)$_{12}$- in the mid-section of the molecule is designed to drive self-organization (constraint 1) as is well known from extensive literature reports [19]. The COOH group in the lower terminus of the molecule in Fig. 1 can provide the necessary attachment to the substrate (constraint 2) for cases of a variety of inorganic oxide materials such as aluminum oxide [19,20]. The single unpaired electron spin appears because of non-compensated valence bond in the β-diketone region (atoms O(2), O(11), and the neighboring carbon atoms near the top of the Fig. 1). This unit of the molecule fits constraints 3 and 4.

We obtained optimal molecular geometry using the UB3LYP/6-311G** method and subsequently calculated atomic spin densities with the UB3LYP/EPR-II model and Eq. (6). Table 1 shows the spatial distribution of the spin densities for the atoms near the radical, as calculated using a Mulliken-type electron population analysis [2]. One can see that the spin density is localized on the two oxygen atoms O(2) and O(11) because each O atom possesses one non-compensated bond with the neighboring C atom while quantum resonance of these two non-compensated bonds leads to sharing one unpaired spin on the two O atoms. The distance between O(2) and o(11) atoms is 0.22 nm. Thus, two oxygen atoms carry the effective electron spin S=1/2.

Table 1. Total atomic spin densities of first molecule as calculated by the UB3LYP/EPR-II model and Eq. (6)

| No of Atom | Spin densities |
| --- | --- |
| 1 C | -0.039422 |
| 2 O | 0.496910 |
| 3 C | 0.054653 |
| 4 C | 0.060413 |
| 5 H | 0.004084 |
| 6 H | -0.003086 |
| 7 H | -0.003086 |
| 8 C | -0.032485 |
| 9 H | -0.003782 |
| 10 C | 0.054497 |
| 11 O | 0.410509 |
| 12 C | 0.001611 |

Using Eq. (7), the overlap population for both the O(2)-C(1) and O(11)-C(8) bonds was found to be approximately 0.57, which is close to the regular C-O bond order. As an example, the value of A for O(46)-C(45) bond is 0.65. Thus, the strong non-compensated valence bond in the β-diketone structure provides the stability of the unpaired spin in the neutral radical molecule.

To explore the effects of local group substitution we substituted the H for a methyl ($CH_3$) group at the carbon between the two carbonyl (C=O) units while leaving the unpaired spin in place, as shown in Fig.2.

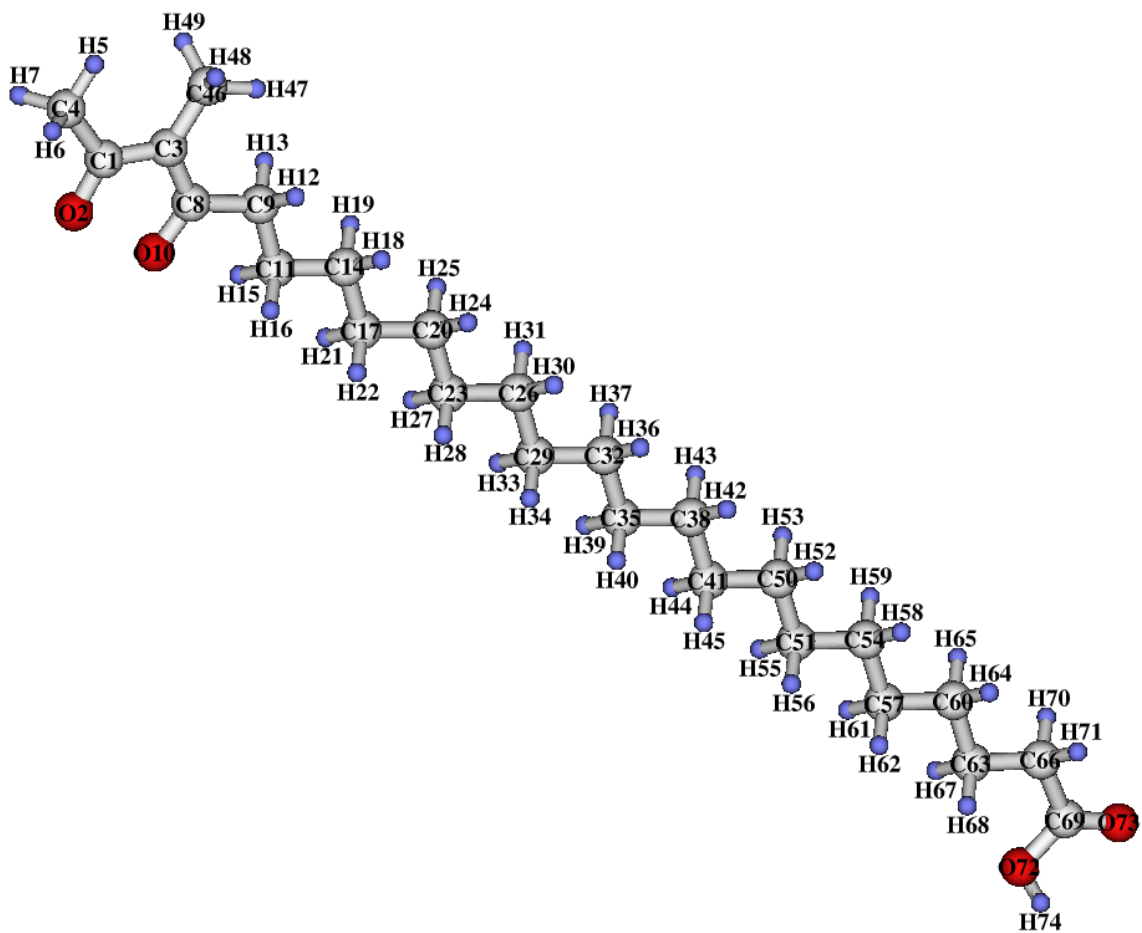

Figure 2. Geometry of a second optimized neutral radical molecule with β-diketone structure.

The results do not change essentially from those of the first molecule. The unpaired spin is localized mostly on the oxygen atoms O(2) and O(10). The corresponding spin densities are 0.49 and 0.42, the distance between those oxygen atoms is 0.215 nm. The overlap population for O(2)-C(1) and O(10)-C(8) bonds is found approximately to be the same, A=0.50. This value of A is slightly smaller than the corresponding value of the first molecule.

We also find that not every neutral radical scan atisfy the four conditions formulated earlier. For example, Figure 3 shows a structure that represents the neutral TEMPO (2,2,6,6-tetramethyl-1-piperidinyloxy) radical molecule.

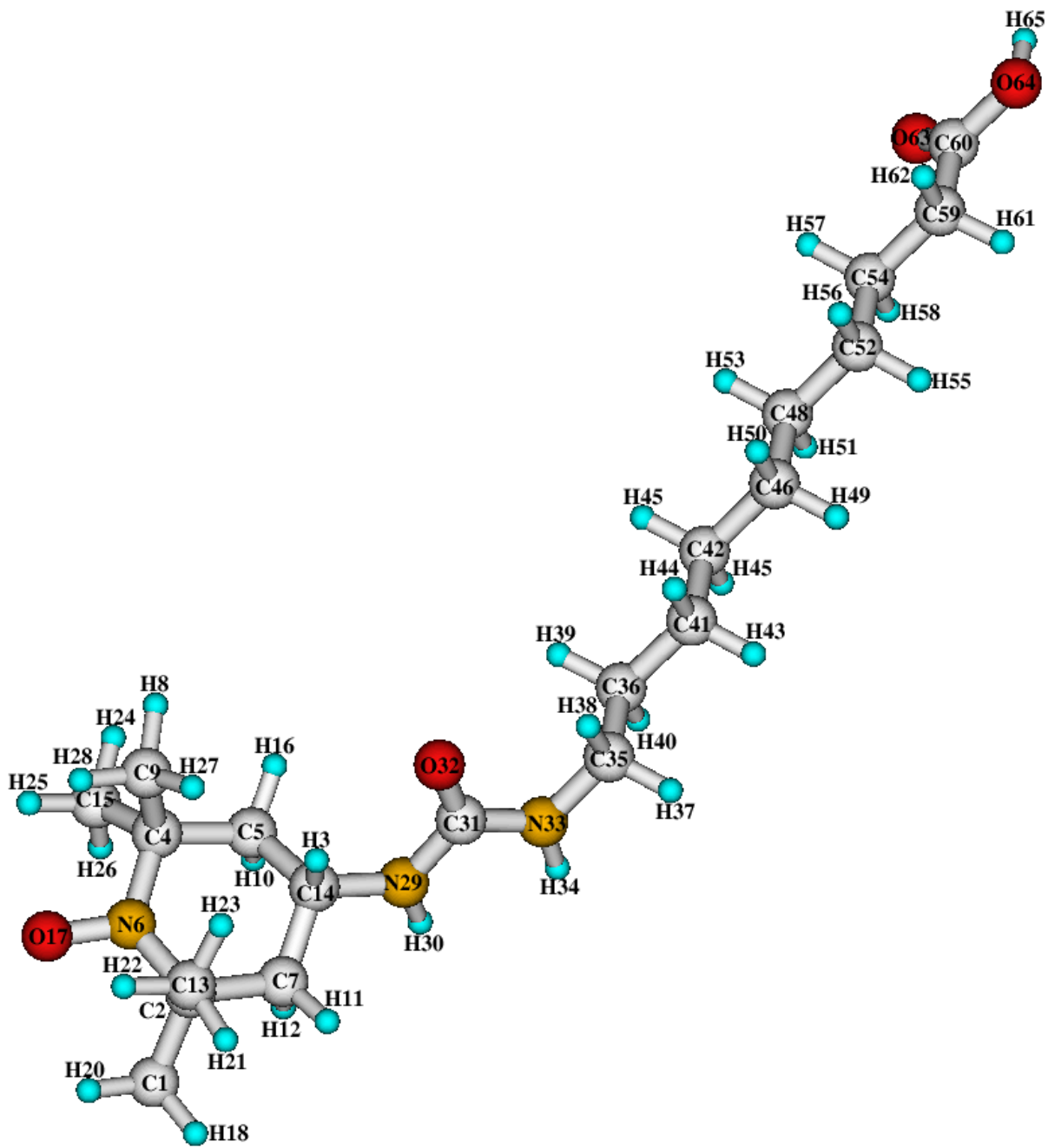

Figure 3. Geometry of optimized neutral TEMPO radical molecule.

For this molecule the unpaired spin is localized mainly on nitrogen (N(6)) and oxygen (O(17)) atoms with spin densities 0.46 and 0.51, respectively. The distance between those atoms is 0.132 nm. Thus, the area of spin localization in this radical molecule is smaller than the corresponding area in the molecules possessing a β-diketone structure. However, the overlap population for non-compensated bonds is the following: A=0.58 for N(6)-C(4), A=0.62 for N(6)-C(2), and A=0.067 for N(6)-O(17). The small value of A for N(6)-O(17) bond indicates the low stability of the unpaired spin which violates the third condition.

Finally, Fig. 4 shows a structure that represents the neutral radical molecule, containing the -(CH$_2$)$_{11}$-CH$_3$ ordering group and non-compensated carbon-oxygen bonds.

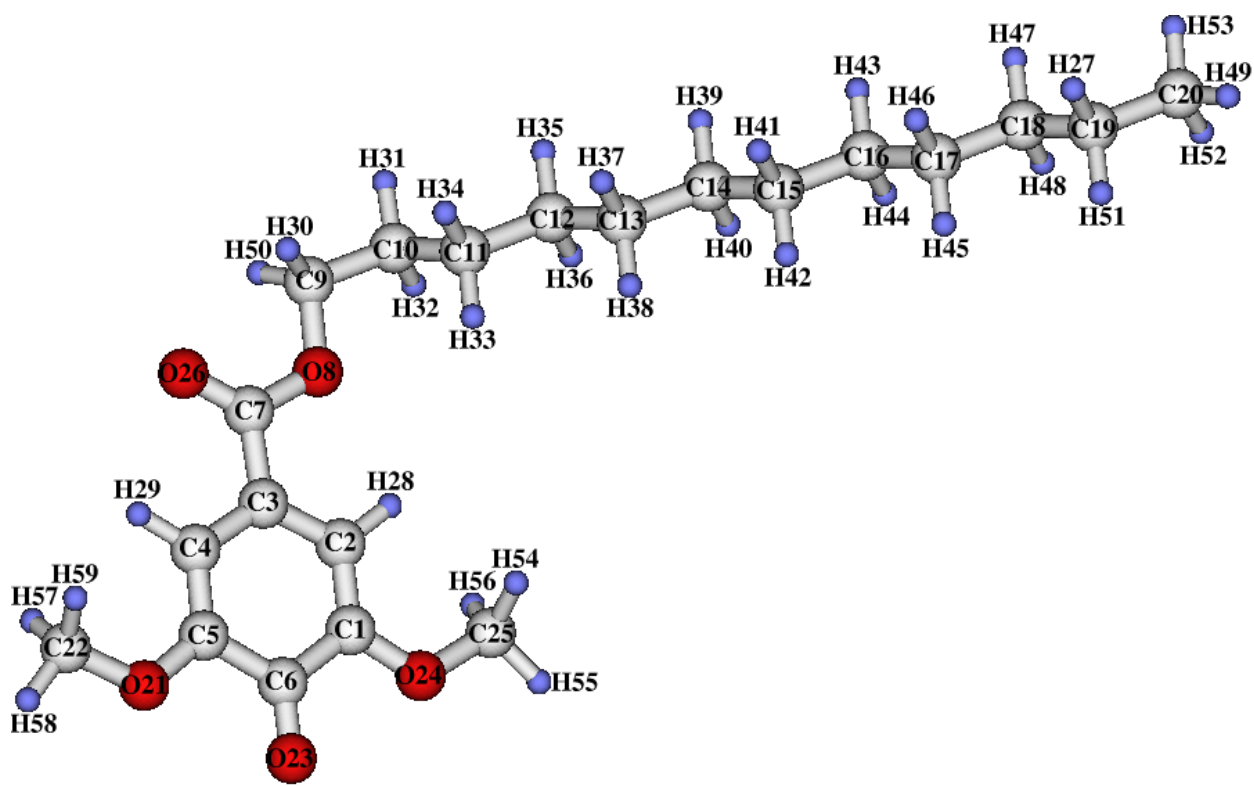

Figure 4. Geometry of an optimized neutral radical molecule containing five oxygen atoms near a phenyl fragment

Our calculations reveal the complicated "antiferomagnetic structure" of the spin density shown in Table 2.

Table 2. Total atomic spin densities of a radical molecule containing five oxygen atoms (Fig. 4) calculated by the UB3LYP/EPR-II model

| No of Atom | Spin densities |
|---|---|
| 1 C | 0.225298 |

| | | |
|---|---|---|
| 2 | C | -0.122716 |
| 3 | C | 0.306939 |
| 4 | C | -0.121119 |
| 5 | C | 0.220468 |
| 6 | C | -0.037633 |
| 7 | C | -0.021294 |
| 8 | O | 0.008971 |
| 21 | O | 0.067679 |
| 22 | C | -0.004785 |
| 23 | O | 0.321491 |
| 25 | C | -0.004772 |
| 26 | O | 0.065633 |

The spin is distributed over oxygen O23 and atoms of phenyl ring C3, C5, C1, C4 with a smaller portion in oxygens O26 and O21. Such spin density hardly corresponds to the conception of a localized spin and the third condition is clearly violated.

## Conclusion

Neutral radical molecules are promising candidates for quantum information processing using spin arrays made from organic SAMs. Their implementation requires adherence to four conditions 1) A specific structure element to drive self-organization of the SAM; 2) a specific chemical group to provide attachment to the selected substrate; 3) a localized unpaired electron spin; 4) a strong non-compensated chemical bond, responsible for the unpaired spin. Using quantum chemical methods based on the density functional theory we have studied several neutral radical molecules. Our analysis of the spatial localization of the electron spin density and the bond orders shows that the neutral radicals with the β-diketone structure satisfy the requirements formulated in our work, whereas other radicals (e.g., galvinoxyl radicals) may not be suitable for quantum computations. However, additional analysis is needed to explore the ability of the suitable molecules to create a stable and ordered SAM.

## Acknowledgments

This work was supported by the Department of Energy (DOE) under Contract No. W-7405-ENG-36, by the National Security Agency (NSA), and by the Advanced Research and Development Activity (ARDA).